# Faster super-resolution imaging with auto-correlation two-step deconvolution

**WEISONG ZHAO,**[1] **JIAN LIU,**[1] **CHENQI KONG**[1] **YIXUAN ZHAO**[1] **CHANGLING GUO,**[2] **CHENGUANG LIU,**[1] **XIANGYAN DING,**[3] **XUMIN DING,**[4] **JIUBIN TAN,**[1] AND **HAOYU LI**[1,*]

[1]*Ultra-Precision Optoelectronic Instrument Engineering Center, Harbin Institute of Technology, Harbin 150080, Heilongjiang, China.*
[2]*Currently with the Department of Electronic Journals, The Optical Society, 2010 Massachusetts Avenue NW, Washington, DC 20036, USA.*
[3]*National Key Laboratory of Tunable Laser Technology, Harbin Institute of Technology, Harbin 150080, Heilongjiang, China.*
[4]*Department of Microwave Engineering, Harbin Institute of Technology, Harbin 150001, China.*
[*]*lihaoyu@hit.edu.cn*

**Abstract:** Despite super-resolution fluorescence blinking microscopes break the diffraction limit, the intense phototoxic illumination and long-term image sequences thus far still pose to major challenges in visualizing live-organisms. Here, we proposed a **s**uper-resolution method based on **a**uto-**c**orrelation two-step **d**econvolution (SACD) to enhance the temporal resolution at lower signal intensity levels. Unlike conventional techniques, such as super-resolution optical fluctuation imaging (SOFI) or stochastic optical reconstruction microscopy (STORM), our model allows 16 frames to generate super-resolution images, without noticeable degradation in recording quality. We demonstrate SACD both in simulated predictions and experimental validations, with the resulting spatial resolution of 64 nm and 2 ~ 10 fold speed improvements. The use of low signal-to-noise ratio acquisition of image sequences, our SACD enables fast, background-free, super-resolution imaging that maybe advance implemented as a suitable tool for rapid live-cells imaging.

**Fluorescence microscopy; Superresolution; Deconvolution; Image reconstruction techniques.**

## 1. Introduction

Various of super-resolution approaches, utilizing nonlinear effects and optical modulations in light microscopy, have been developed to have our vision surpass the diffraction limited resolution [1]. The major element to achieve super-resolution imaging is based on the fluorescence bright-dark modulations, either in structured illumination manners, e.g., stimulated emission depletion (STED) [2] and saturated structured illumination microscopy (SSIM) [3], or in fluorescence blinking manners, e.g., stochastic optical reconstruction microscopy (STORM) or photo-activated localization microscopy (PALM) [4–6], and super-resolution optical fluctuation imaging (SOFI) method [7]. The STED or SSIM category required costly optical system and complex alignments. In contrast, the stochastic emission method, such as SOFI and STORM, has relative easy-build microscopy devices, but the temporal resolution still limits the applications of this technique in rapid live-organisms imaging [8–11]. In their studies, they demonstrated that the algorithms allowed approaching the *Cramér-Rao* theoretical precision limit as the distance of the emissions is up to ~1.5 μm [12]. Therefore, the key factor becomes critical to keep the emitting photons activity in each single sensor image of the frame series recorded under the condition boundaries of the analytical approach used [13]. However, it will be challengeable when imaging cellular dynamics or visualizing diverse anatomical traits in living organisms across high spatial-temporal dimensions.

In recent years, a number of numerical algorithms and theoretical models in particular aim to addressing this maximum density limitation based on analysis the artificial or spontaneous fluorescence intensity blinking and blench [14–16]. For example, stochastic optical fluctuation imaging, i.e., SOFI [7], was used to calculate the temporal correlation cumulant. Super-resolution radial fluctuations, i.e., SRRF [17], enables to analyze image radiability distribution. Deconvolution-STORM, i.e., deconSTORM [18], allows averaging extensively deconvolved images. Iterative deconvolution approach, with single frame initial condition, enabling the image contrast and resolution enhancements, is always used in biological imaging to improve the image quality [19].

Lucy-Richardson (LR) [20,21] deconvolution is a Bayesian-based derivation technique resulting in an iterative expectation-maximization (EM) algorithm. LR deconvolution plays an ever more important role as its simplicity and advanced performance, and it is frequently employed as a particularly useful tool for improving the imaging during its post-exposure time stages. According to the model practical applications, the following two conditions appear necessary for LR deconvolution [22]: (i) the number of estimated parameters is much smaller than its observations, (ii) the algorithm is terminated at an early stage of iterative calculation. The first condition is easy to satisfy in fluorescence biological imaging, as its dark background and the sparse character of cell imaging in fluorescent microscopy systems. Meanwhile, the second condition limits the resolution to the single image by use of deconvolution calculation. These algorithms developed could reduce the time resolution and without the requirement of special specimen. In spite of this, hundreds of frames and long acquisition time are still major limitations for generating the super-resolution dataset for rapid imaging.

In this paper, we propose a novel super-resolution method using auto-correlation with two-step LR deconvolution (SACD). Unlike more conventional methods, SACD reduced the number of frames indispensable from hundreds to tens frames when producing super-resolution images. In the model procedures, we first used deconvolution to 16 frames for an extreme short time, and it can be regarded as the filter theoretically. Then the multi-plane auto-correlation (MPAC) was involved to further inhibiting the image artifacts appearing in the background, which was important to the next deconvolution stage. As the background being suppressed, we took the second step deconvolution to the single image after the correlation process. Combining the deconvolution and correlation together, we made the order of correlation low and the times of deconvolution short, which improved the linearity of image intensity, reduced the artifacts of super-resolution image and the required image frames. We demonstrated the proposed SACD both in numerical simulations and experimental datasets, and the resulting temporal resolution has a 2—10 fold speed improvements with the spatial resolution of 64 nm. Our SACD method can be used to analyze all kinds of fluctuation data and a wide range of density for single-molecule localization microscopy (SMLM) data.

This article is structured as follows. In Sec. 2, the basic LR deconvolution approach is briefly reviewed. Following a discussion of auto-correlation process with its calculation method, MPAC method is introduced. In Sec. 3, combining LR deconvolution with MPAC, our SACD model is developed. In Sec. 4, simulations are showed and analyzed. SACD algorithm is implemented to the corresponding experimental raw date, enabling high-speed, background-free, super-resolution imaging. Finally in Sec. 5, a brief conclusion is given.

## 2. Theoretical model

In this Section, we begin by describing the specific calculating flow of LR deconvolution and reviewing the basic theory of MPAC separately, and then show how they can be usefully combined. Under the two-step LR deconvolution with auto-correlation processing, the SACD algorithm is developed with a detailed discussion.

### 2.1 Theory of LR deconvolution

In the LR deconvolution approach, it assumes that an observed image can be described approximately by the *Poisson count* model [20,21]. The probability distribution of the light intensity is given as the follows:

$$P(g \mid f) = \prod_{r \in R} \frac{[f(r) \otimes h(r)]^{g(r)} e^{-[f(r) \otimes h(r)]}}{g(r)!} \tag{1}$$

where $\otimes$ denotes the convolution operation, $h$ represents the *Point Spread Function* (PSF) of the optical system, $f$ is the object which is our expectation, $R$ is the total set of image pixels of observed image $g$, and $r$ denotes a 2-D discrete coordinate.

As LR deconvolution based on the max likehood estimation (MLE), the posterior probability of the object can be described by Eq. (1). With the value of probability getting higher, the object estimated will be closer to the true object. In this step, taking the '-log' operator into Eq. (1), we define that $E(f) = -\log(P(g|f))$ and then it can be rewritten as:

$$E(f) = \sum_{r \in R} f(r) \otimes h(r) - g(r) \log[f(r) \otimes h(r)]. \tag{2}$$

Since the functional $E(f)$ is convex in $x$, it is easily to solve the convex function with EM algorithm and the expression is given by:

$$f_{k+1}(r) = f_k(r) \left( h(-r) \otimes \frac{g(r)}{h(r) \otimes f_k(r)} \right) \equiv \psi(f_k) \tag{3}$$

In Eq. (3), $k$ is the number of iterations, and $f_k(r)$ is the estimated image at iteration $k$. Here $\psi(...)$ is called LR function in LR deconvolution algorithm.

Based on Eq. (3), we further introduce the acceleration algorithm as proposed in previous studies [23] to accelerate the LR deconvolution, and after this, the calculation forms can be rewritten as a serious of equations as follows：

$$f_k = x_k + \alpha_k h_k, \tag{4a}$$

$$h_k \equiv x_k - x_{k-1}, \tag{4b}$$

$$x_{k+1} = f_k + g_k, \tag{4c}$$

$$g_k \equiv \psi(f_k) - f_k, \tag{4d}$$

$$\alpha_k = \frac{\sum g_{k-1} \cdot g_{k-2}}{\sum g_{k-2} \cdot g_{k-2}}, \ 0 < \alpha_k < 1. \tag{4e}$$

It is notable that Eq. (4) is the acceleration of this algorithm, and is achieved without the need for the cost function which proposed to be determined or minimized. Therefore the algorithm is more stable and the acceleration factor can be derived by experimental results directly, without the assignment during the post-processing.

According to the above discussion, the PSF of the conventional microscopy system can be calculated from the expression followed [24]:

$$h(r) = \left| \int_0^1 J_0(k\sin(\alpha)\rho r) \exp(-jkz\sin^2(\alpha/2)\rho^2) \rho d\rho \right|^2 \quad (5)$$

where $\sin(\alpha)$ represents the numerical aperture (NA) of the microscope system, and $k$ is the wave number of the emitting fluorophore.

## 2.2 Theory of MPAC

Prior to presenting the theory of multi-plane auto-correlation (MPAC), we first introduce a concept of point spread function (PSF) of the microscopy imaging system to characterize its optical performance and response. For example, when recording an image using a wide-field microscopy, a single spot emitting light and propagating to the native imaging plane, in which a detector (typically a CCD camera) is placed, and the captured light intensity distribution in the final imaging plane is defined as the PSF of the microscope system.

Before the calculation, we give the mathematical expression of PSF in Eq. (6), in which the value of PSF can be fully determined by the optical properties of the optical system, i.e., object lens NA and wavelength $\lambda$. It is worth noting that the spatial resolution of one microscopy depends on the measured size of its PSF, which is defined as its full-width-half-maximum (FWHM). It is well known that, in STORM/PLAM system [4–6], such critical problem was circumvented by reducing the dense of light fluorescent to estimate each location of emitter. As to the case of SOFI scheme, it does not require such sparsity molecule emissions through calculating high order cumulant of correlation to raise the PSF to $n$th order.

As similar to SOFI system, here we calculate the MPAC of the image sequences which have been deconvolved in previous steps to further reduce its background, noise and improve its spatial resolution by narrowing the width of PSF. To achieve this, we consider a sample that is labeled with $N$ fluorescing molecules at positions, and the brightness $s_j(t)$ is time-dependent, in which the range of intensity is $1 \leq k \leq N$. To do so, the time varying image, which is being recorded at any moment in real time $t$, can be given by the sum of fluorescent emitting light intensity:

$$F(\mathbf{r},t) = \sum_{k=1}^{N} U(\mathbf{r} - \mathbf{r}_k) \cdot \varepsilon_k \cdot s_k(t) \quad (6)$$

where $\varepsilon_j$ is the maximum brightness of the $j$th molecule, and $s_j(t)$ describes its temporal intensity fluctuations.

It is assumed that the emissions of different fluorescent molecules are not correlated in time and the samples are in stationary equilibrium during acquisition, then we can get its auto-correlation function as the expression:

$$\begin{aligned} C(\mathbf{r},\Delta t) &= \langle \delta F(\mathbf{r},t+\Delta t) \cdot \delta F(\mathbf{r},t) \rangle_t \\ &= \sum_k U^2(\mathbf{r}-\mathbf{r}_k) \cdot \varepsilon_k^2 \cdot \langle \delta s(t+\Delta t) \cdot \delta s(t) \rangle \end{aligned} \quad (7)$$

In Eq. (7), we defined the identical relation that $\delta F(\mathbf{r},\theta) \equiv \sum_k U(\mathbf{r}-\mathbf{r}_k) \cdot \varepsilon_k (s_k(\theta) - \langle s_k(\theta) \rangle_\theta)$, and it also should be noticed that the PSF is replaced by a distribution, and its value can be calculated as the square of the original PSF.

In this part, it is worth to emphasize that, as the background and noise information of the original image sequences are not correlated in our model calculation, these values are tended to zero in the reconstructed image. At this point, the processed image sequences are suit to the second step of deconvolution. It is noticeable that, in this article, we only involved about tens frames into the model calculation, which means it should have a little intensity discontinuity and much statistical error in the image after auto-correlation for such few images to analyze. Therefore, according to these considerations, we chose the multi-plane method to compromise this issue, and the full process of MPAC is as shown in Fig.1.

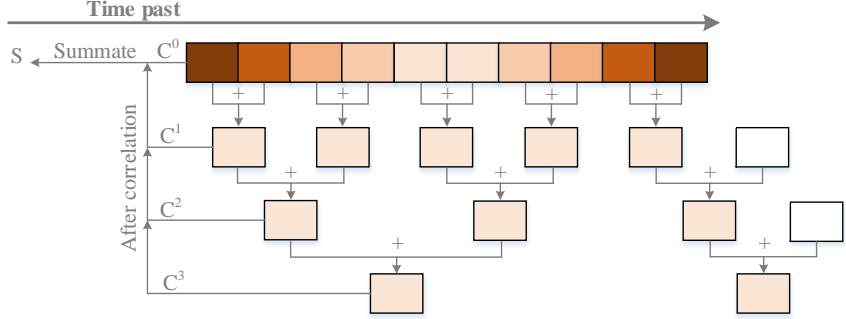

Fig. 1. Flow chart illustrating the primary mechanisms of MPAC.

At the last step of MPAC, we generate a few stacks of images by binning every two subsequent frames, from previous stack, into a single frame. Then we calculate auto-correlation to these stacks respectively and the final reconstruct image can be obtained as a simple sum as following expression:

$$S(\mathbf{r}) = \sum_k C^k(\mathbf{r}) \qquad (8)$$

In Eq. (8), the $S(\mathbf{r})$ denotes the MPAC result, and the $C^k(\mathbf{r})$ presents the auto-correlation result in the $k$th plane of the input frames.

## 3. SACD Method

According to the detailed discussions on two above algorithms, i.e., LR deconvolution and MPAC, in this section, we now develop a novel super-resolution imaging method based on auto-correlation and two-step deconvolution, i.e., SACD, to improve the imaging speed, under the initial conditions of low signal intensity levels and fewer input frames. Based on the mathematical derivations, the LR deconvolution and the MPAC can be combined and then involved in our proposed SACD method to achieve super resolution imaging. The specific flow chart and corresponding simulated predictions of our proposed SACD model are presented in Fig.2.

In the first step of SACD method, we acquired a short diffraction limit image sequence of $I_m$ ($m = 1 \ldots M$), and then put these frames into the proposed SACD calculation process. With these input images, $g_m$, and considering with the basic parameter of the optical imaging system, such as, $N.A.$, $\lambda$, the pixel size of CCD, and the up-sample factor, the SACD model starts to analysis and process the original images. Here it is worth to emphasize that in order to get smaller pixel size and larger pixel number, in the meanwhile keeping the real image content unchanged, we chose the Fourier interpolation [25] to match the improved spatial resolution.

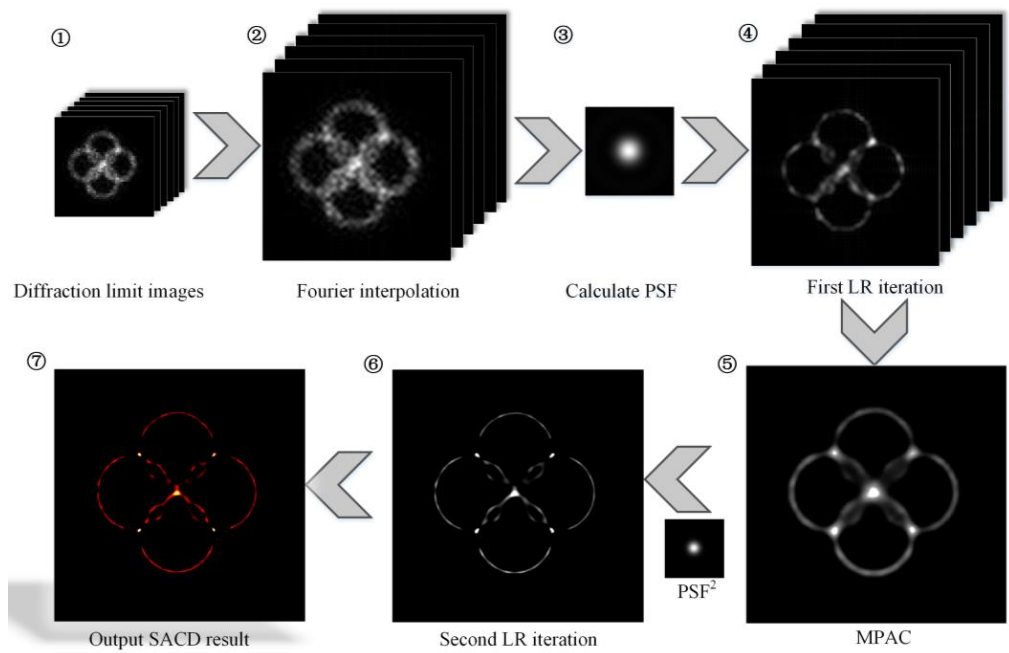

Fig. 2. Flow chart illustrating the main steps of SACD model. Step 1. acquisition of short diffraction limit image sequence, i.e., $I_m$, $m = 1, 2, \cdots, M$; Step 2. Fourier interpolation for $I_{1...M}$, with magnification times (mag.) to $g_{1...M}$; Step 3. calculating the PSF with the obtained parameter values of optical system; Step 4. the first 10 LR iterations is carried out with the calculated PSF; Step 5. analyzing these processed image sequence using MPAC method; Step 6. the second LR iteration of $n$ times is carried out with the square of PSF ($PSF^2$). Step 7. SACD super-resolution results are extracted with the pre-set magnification times.

It is clear that in most super resolution imaging approaches, i.e., SOFI, SRRF and STORM/PLAM, a prior problem to be resolved is how to reduce the pixel size. The main motivation is to have the acquired pixels in the CCD plane match the improved resolution in the produced super resolution images. In conventional methods [4–7], it can be achieved by inserting an additional lens to increase the magnification of the imaging system. However, in general, this treatment would not be enough to match the final resolution of the images.

According to the above analysis, in the second step, we introduce an algorithm of Fourier interpolation, which is an easy and accurate pre-treatment, aiming to process the obtained images. By including Fourier interpolation in our SACD model, it will figure out this ticklish problem directly, and make the pixel size match the improved resolution more accurately in an easy manner.

Based on the basic theory of *Optical Transfer Function* (OTF), it is known that the imaging system has only a finite support [26–28], which means that the high frequency area of Fourier transform of diffraction limit images would drop at zero. By padding the Fourier-transformed image with zeros and Fourier transforming, we obtain the resulting image with smaller pixel size and larger pixel number. It should be noticed that in our model, the Fourier interpolation would not change the frequency information and provide the artifacts-free for the processed images with a better signal-to-noise (SNR). After the second step, we can obtain the interpolated images, i.e., $g_m$. Since the pixel size of the interpolated image has been processed smaller, such pixel size being used to calculate PSF should be regarded as pixel size dividing mag.

Next, by use of the SACD model setting parameter values to generate PSF numerically, the short step deconvolution, i.e., first 10 LR iterations, is implemented to the interpolated image sequence. Here it is worth emphasizing that if the unknown object extends up to the boundary of the image domain, consequently the effect of the PSF is to generate an image which is not completely contained in the field-of-view of the microscope. Then the FFT-based deconvolution result can produce annoying boundary artifacts. Therefore we padding the image before fast Fourier transform algorithm (FFT) with zeros [29] to reduce this annoying boundary. Moreover it is necessary to point out that the values of PSF distribution should be normalized as its summed values of 1.

In some previous studies [30–32], common denoise methods focus on smoothing the image information. To do so, it will change the high frequency information and make the image dataset unreliable, which will further influence the subsequential steps. Nevertheless, the LR deconvolution takes account of the *Poisson* noise characteristic of the photon counting process in cameras, and it will filtrate the noise. As the short step of LR deconvolution, it can sharp the image without artifacts production.

Due to only 10 iterative deconvolutions applied on each frame, the MPAC is used to analyze the image sequence, in the fifth step of SACD, as illustrated in Fig. 2. Considering that the reconstructed MPAC is background-free, the deconvolution is demonstrated as a very suitable manner to improve the image resolution. If there is representing very rare of imaging background, the deconvolution algorithms would be tending to operate better performance. This corresponding conclusion has also been refereed and discussed as the first necessary conditions during the above introduction of LR deconvolution.

At the final step, long step LR deconvolution is used on the single image after the MPAC reconstruction with the squared PSF. The proposed SACD super-resolution results are obtained with the pre-set magnification times.

## 4. Results

To demonstrate the availability and accuracy of our proposed SACD method, we applied it to process different image sequences, including the standard example data, the fluctuation data, and high density single-molecule localization microscopy (SMLM) data. It shows that SACD technique exhibits considerable good performance on analyzing all these three types of datasets. Compared to the other super-resolution methods, SACD has its unique advantages that this method needs less frames and remains achieving the similar resolution as the conventional approaches.

### 4.1 Simulated results

The simulated image stack can be generated by assuming *Poisson* blinking model [33], where the length of switching times *t*-on and *t*-off are calculated using a *Poisson* distribution with the average on time value of $\tau_{on}$ = 3 ms and the average off time value of $\tau_{off}$ = 7 ms. As can be observed in Fig.3, the example object consists of four circles, with the radius of 500 nm. In each circle there are 180 emitters uniformly located along the circumference of the circle. For practical purposes, there also has a 5 percent *Poisson* noise being added to simulate an experimental fluorescence imaging condition. The image stack in this work contains 16 frames, and an emission wavelength of $\lambda$ = 488 nm, with the numerical aperture NA = 1.49 and the pixel size of 65 nm.

Figs. 3(a)-(c) the resulting images are reconstructed by using 16 frames, with the methods of mean wide field image, MPAC image without two-step deconvolution, and the SACD reconstructed image, respectively. In the SACD reconstruction model, the number of magnification times is 8 and the iterations of second step deconvolution is chosen of 20. The intensity profiles of the projection through the dotted lines in Figs. 3(a)-3(c) are quantified and shown in Fig. 3(f). As being to closer to the intersection of circles, it is harder to distinguish the structure.

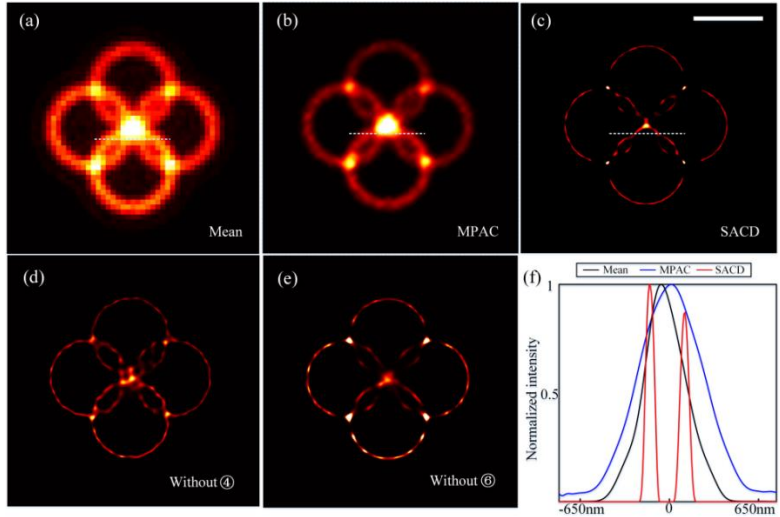

Fig.3. Simulation results of the circle structure sample. (a) Mean wide-field image, (b) MPAC image, (c) SACD reconstructed image, (d) SACD reconstructed image without first deconvolution step, (e) SACD reconstructed image without second deconvolution step. (f) Profiles of the projection through the dotted lines in (a), (b) and (c). Scale bar 1 μm.

To demonstrate how the evolution of SACD algorithm is related to the auto-correlation and the two-step LR deconvolution in more detailed, in Figs. 3(d)-(e), we presents the SACD reconstructed image without the first and second deconvolution step, respectively. When taking out the first deconvolution step, as seen in Fig. 3(d), the center of reconstructed image can be seen full of artifacts, especially around the profiles of four circles. Due to lack of input frames recovering the image, the statistic noise of auto-correlation is hard to ignore. In addition, the auto-correlation could not release all the noise under such condition of insufficient frames. As to the second deconvolution step loss, observing in Fig. 3(e), after the first step deconvolution and MPAC processing, the reconstructed image obtained is less noise and almost background free. By doing this, the resulting image will be suitable to the continual deconvolution step to improve its resolution sequentially.

Examining Fig. 3(f), it can be clearly seen that, only our SACD model can distinguish the two lines near the intersection. Therefore, this result, to some extent, can fully prove the ability of our SACD model improved the resolution of the optical imaging system. Meanwhile, as the simulated structure used in Fig. 3 is standard example, in which the true structure has been already known, the model predictions can also demonstrate the accuracy of our SACD model.

### 4.2 Fluctuation data

In this part, we demonstrate the experimental results by super-resolution radial fluctuations (SRRF) technique and our SACD model, respectively, and the reconstructions between SACD and SRRF using live-cell imaging data are compared. The SRRF method was proposed by Ricardo Henriques et al. [17], which is comprised of spatial and temporal parts. After calculating the radiality map of each image, they involved the temporal analysis such as auto-cumulants. In this paper, the SRRF reconstructed image is acquired from RHenriques-Lab online data source [34]. The fluctuation live-cell image data of mEGFP-Microtubules in HeLa H2B-mEGFP-α-tubulin stable cell line is provided. The 200 frames data was acquire with 100 f.p.s., NA = 1.49, and a pixel size of 107 nm.

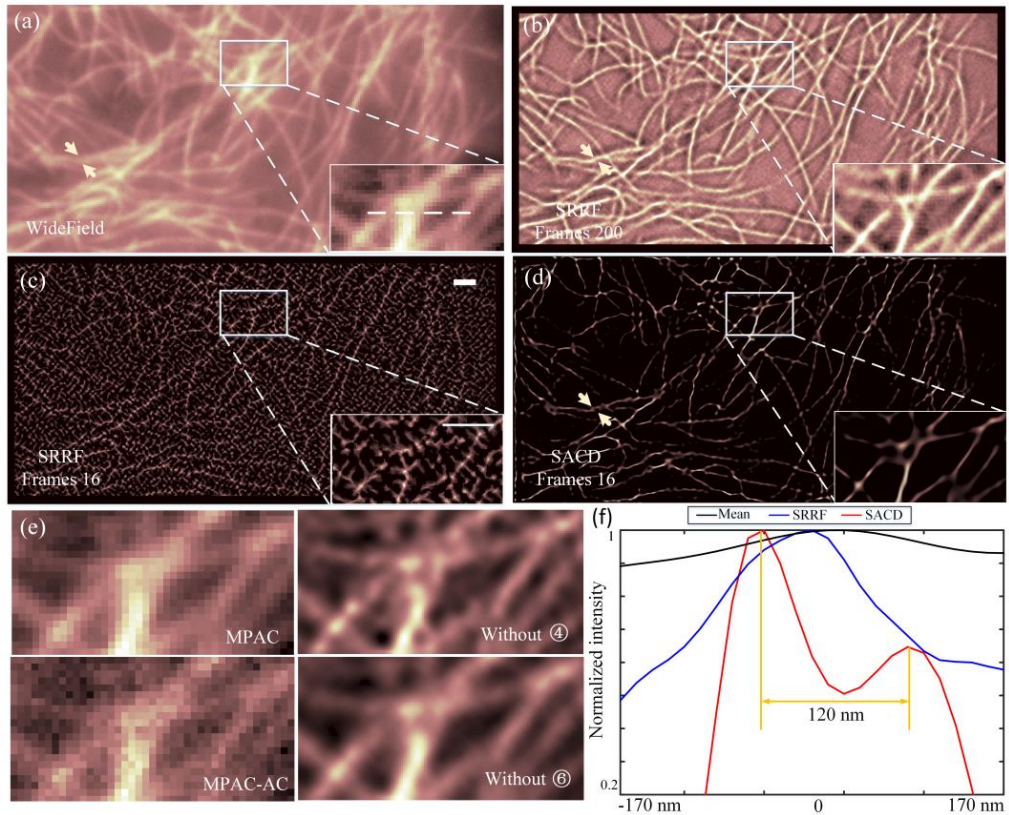

Fig.4. SACD result of fluctuation images and comparison between SACD and SRRF using live-cell imaging data. (a) Mean wide-field image, (b) SRRF reconstructed image with 200 frames, (c) SRRF reconstructed image with 16 frames SACD reconstructed image, (d) SACD reconstructed image, (e) Right: MPAC and error map between MPAC and AC image; left: SACD reconstructed image without first and second deconvolution step. (f) Normalized line profiles taken from the regions between the yellow arrowheads for corresponding images in (a), (b) and (d) showing separated features. Scale bar 1.5 μm. AC: auto-correlation.

In Fig. 4(a) and (b), we show the mean of fluctuation image sequence and the SRRF super-resolution results, respectively. Applying the SRRF *ImagJ* [35] *plugin* [34], we acquire the super-resolution result with 16 frames of dataset, which is shown in Fig. 4(c). With insufficient frames involved in the analyzing model, the deviation of SRRF result is not precise enough, and therefore the imaging quality of SRRF reconstructed result is not satisfactory. In general, the SRRF method needs at least 200 frames to reduce the stochastic error and enhance the contrast between background and sample signal distribution. However, the image reconstructed by our SACD model seems more desirable, in which there is only with 16 frames included, as can be seen in Fig. 4(d). Zooming in the images of the white boxed region on the lower right corner of Fig. 4(a)-(d), we show the expanded views of the microtubule distributions in the cell. We can observe that SACD with 16 frames can get the similar imaging capacity to the SRRF result, in which the image is reconstructed by involving 200 frames. Moreover, it is also worth noting that there is even less background appearing in our SACD model results. For all of these demonstrations, it is clear that SACD method could enhance the time resolution of the super-resolution reconstruction due to less frames input.

For a better explanation, in Fig. 4(e), we further demonstrate the results of MPAC and error map between multi-plane auto-correlation (MPAC) and auto-correlation (AC), without the first deconvolution step and the second deconvolution step, respectively. As in the left side of Fig. 4(e), with only 16 frames AC calculating, the result contains stochastic error in the reconstructed image. Nevertheless, in the case of MPAC algorithm, it restrains the error for some degree to help the second step deconvolution reducing the artifacts. In the right side of Fig. 4(e), we can observe that without the first step deconvolution, after MPAC and the second deconvolution processing, the reconstructed result is full of artifacts. It is also observed that without the second step deconvolution, the resolution of the image is higher for super-resolution imaging with less artifacts production.

The intensity profiles of the projection between the yellow arrowheads in Figs. 4(a), (b) and (d) are presented in Fig. 4(f). The line of mean image, SRRF and SACD respectively proves that the full-width-half-maximum (FWHM) of SACD is much sharper than the other methods, and moreover the background free appears in the case of SACD reconstruction. The mean resolution of SACD could reach about 120 nm and meanwhile the temporal resolution of SACD is improved from 0.5 f.p.s to 6.25 f.p.s.. Here, it should be noticed that, with so few frames to analyze, and the high axial resolution of SACD, the SACD result may filtration some defocused sample structure compared to SRRF result. For a better structure observation in Fig. 4, we involved the colorbar 'pink' rather than the common used 'hot'.

### 4.3 High density SMLM data

According to the mechanism of SACD, the proposed our algorithm can also recover super-resolution image using the detected image sequence of the high-density STORM. To verify this viewpoint, 500 frames (only use the first 256 frames) of STORM raw images obtained from the EPFL website [36] are used. The initial parameter values in this STORM experiment are given as follows: NA = 1.3, frame rate is 25 f.p.s., and the effective CCD pixel size is 100 nm.

As presented in Fig. 5(a)-(b), the first 256 frames and 16 frames with a higher density obtained by adding each 16 frame of the recorded images together, are used to the model analyzation, respectively. In addition, we compare three typical methods in detailed and present the results in Fig. 5(a)-(b), i.e., SACD (magnification times 8 with 40 iterations), bSOFI (magnification times 4 with the order 4), and Thunder-STORM [37] (magnification times 8). As Thunder-STORM collected methods extensively, it performs almost best compared to the other SMLM software packages, so we chose Thunder-STORM to demonstrate the effectiveness of SACD.

In Fig. 5(b), it is noticed that, through the high density STORM data, the photons distribution is much sparse in the 16 frames case, and such input signal is not enough to display the whole sample information. In spite of incomplete reconstructions, our SACD still performs well on the accurate photon locating. This result proves that SACD will be more befitting the live-cell imaging as less frames required and higher density fluorescence response (lower illumination intensity) [38].

Besides the Thunder-STORM and bSOFI [39], we also demonstrate the SRRF and ESI [40] reconstructed results in Fig. 5(c). Even 256 frames being included, to some extent, it is not abundant for these conventional super-resolution methods. For the Thunder-STORM case, it is unable to distinguish structure with high fluorescence density or more complex distributions. Similarly, the resolution and some recorded areas in bSOFI results are not showing good imaging qualities. In comparison, our SACD resolved these problems and the resulting tubulins are recovered more continuity with the artifact-minimized.

As can be examined in Fig. 5(d), the biological structured line reconstructed by our SACD model attains the resolution about 64 nm, which is very similar to the Thunder-STORM and better than the bSOFI scheme. Moreover, we note that images reconstructed by SACD model are seems more continuous, with higher contrast and less recording 'error'.

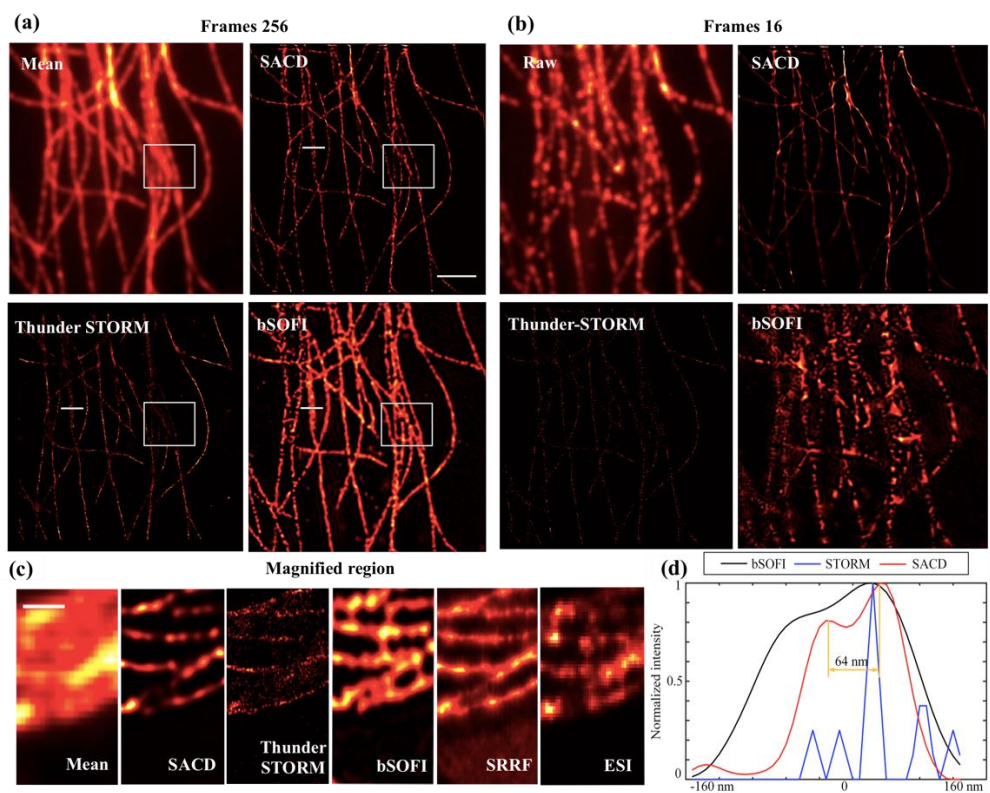

Fig.5. SACD result of high density STORM images and comparison with the other super-resolution methods. (a)-(b) The meaned wide-field image and the reconstructed images: 256 and 16 frames. (c) The magnified result of white boxed region. (d) Profiles of the projection through the dotted lines in (a). Scale bar (a) 2 μm; (c) 500 nm.

## 5. Conclusion

In this paper, we proposed a new super-resolution method based on fluorescence blinking and blench microscope systems. Utilizing LR deconvolution as two separated steps and combining with the multi-plane auto-correlation (MPAC), SACD model is developed. In this SACD model, the time-varying fluorescence intensity images can be analyzed and produce super-resolution images under insufficient frames involved or the sparse photons level. By using of three typical datasets, i.e., the simulated fluctuation data, the experimental fluctuation data, and high density SMLM data, we demonstrate the accuracy and the effectiveness of our proposed SACD model. Our SACD and more conventional methods are also compared in detailed and the reconstructions substantially outperforming these current algorithms.

This SACD model is validated by performing a set of reconstructed images under the short simulated and experimental image sequence. Under such 10%—50% less frames conditions, the more conventional super-resolution methods present limited performances. Even under such insufficient frames input, leading a 2－10 fold speed improvement of imaging, it still attains the resolution to up to 64 nm, with the unnoticeable degradation in image quality. Therefore, our SACD model provides another promising option for biological live-cell imaging as it overcomes the primary challenge of less frames or lower illumination power when imaging of rapid live-organisms.

The SRRF, Thunder-STORM, ESI results were generated with *ImageJ plugi*, bSOFI and SACD were generated with Matlab 2017b. The source code written in Matlab environment is available of download online: https://github.com/WeisongZhao/SACD. As SACD requires less frames and performances well in wide range of datasets, we hope our SACD model including related source codes and supplementary could help for the fast live-cell super-resolution imaging research.

Much work remains to be done. As our SACD model is based on auto-correlation and deconvolution, which can produce background-free super-resolution images. The SACD has the ability in further 3D imaging. On the other hand, it also should be mentioned that SACD has its limit for the signal-to-noise (SNR) of the image sequence. For the SNR is lower to a certain extent, such as: less exposure and long term imaging, the SACD reconstructed result should suffer from artifacts. This limit could be resolved by increasing the number of image sequence. To keep the time resolution of SACD with low SNR will be our primary working focus in the future.

## Funding

This work was supported by the Fundamental Research Funds for the Central Universities (AUGA5710050218).

## Acknowledgment

Correspondence and requests should be addressed to Haoyu Li (lihaoyu@hit.edu.cn).